# Quantum-confined single photon emission at room temperature from SiC tetrapods

Stefania Castelletto[a], Zoltán Bodrog[b], Andrew P. Magyar[c], Angus Gentle[d] Adam Gali[b,e], and Igor Aharonovich[d,*]

[a] School of Aerospace, Mechanical and Manufacturing Engineering RMIT University, Melbourne, Victoria 3000, Australia;

[b] Institute for Solid State Physics and Optics, Wigner Research Centre for Physics, Hungarian Academy of Sciences, P.O.B. 49, H-1525, Budapest, Hungary;

[c] Center for Nanoscale Systems, Harvard University, Cambridge MA 02138 USA;

[d] School of Physics and Advanced Materials, University of Technology Sydney, Ultimo, NSW 2007, Australia;

[e] Department of Atomic Physics, Budapest University of Technology and Economics, Budafokiút 8. H-1111, Budapest, Hungary

* Correspondence should be addressed to Igor Aharonovich, igor.aharonovich@uts.edu.au

**Abstract**

Controlled engineering of isolated solid state quantum systems is one of the most prominent goals in modern nanotechnology. In this letter we demonstrate a previously unknown quantum system namely silicon carbide tetrapods. The tetrapods have a cubic polytype core (3C) and hexagonal polytype legs (4H) – a geometry that creates a spontaneous polarization within a single tetrapod. Modeling of the tetrapod structures predict that a bound exciton should exist at the 3C-4H interface. The simulations are confirmed by the observation of fully polarized and narrowband single photon emission from the tetrapods at room temperature. The single photon emission provides important insights towards understanding the quantum confinement effects in non-spherical nanostructures. Our results pave the way to a new class of crystal phase nanomaterials that exhibit single photon emission at room temperature and therefore are suitable for sensing, quantum information and nanophotonics.

## Introduction

Fluorescent nanostructures that can emit single photons on demand are important for a variety of applications spanning biosensing, photocatalysis, photovoltaic and quantum technologies[1-7]. A new class of materials that recently attracts considerable attention is the homogeneous heterostructure - alternatively termed crystal phase quantum dots[8]. These nanomaterials have identical chemical composition with alternating polytypes – for instance wurtzite (WZ) and zincblende (ZB) in the arsenide family[9-11]. Indeed, engineering InP nanowires with periodic WZ/ZB yielded great insights into the understanding of exciton dynamics and realization of quantum dots embedded within nanowires[12] and lasing[13].

Silicon carbide (SiC) is another example of a material with over 200 known polytypes, with the cubic 3C and hexagonal 4H structures among the most common polytypes. SiC was studied for many decades for its outstanding mechanical, electronic and thermal properties, with a large variety of applications including light emitting diodes (LEDs)[14] and micro- and nano-electromechanical systems[15]. Recently, coherent control over spin defects in different polytypes of SiC[16, 17] and the realization of single photon emission from bulk 4H SiC was demonstrated[7] underpinning its prime role in integrated multifunctional quantum devices.

In this paper we report that silicon carbide tetrapods can be harnessed as room temperature, single photon emitters due to the quantum confinement effect at their structure. The origin of the quantum confinement and the non-classical emission is the homogeneous heterostructure of the 3C core and 4H legs that form a crystal phase SiC tetrapod. Earlier works on ZnO tetrapods and CdTe tetrapods have also shown that a single tetrapod can have a zinc-blende core and wurtzite arms.[18-22]

## Experimental

The tetrapods were grown using microwave plasma chemical vapor deposition (CVD) from adamantane molecules embedded in a solgel matrix. The growth conditions were: microwave plasma power of 900W under 99 % methane and 1 % hydrogen, with no external heating source. At these conditions, the typical substrate temperature is 950°C. The structure of a SiC tetrapod is shown schematically in Figure 1a. Figure 1b shows a high resolution scanning electron microscope (SEM) image of the tetrapods. The majority of the tetrapods have an average leg length of 100 nm and leg diameter smaller than 50 nm, with a core as small as several nanometers[23].

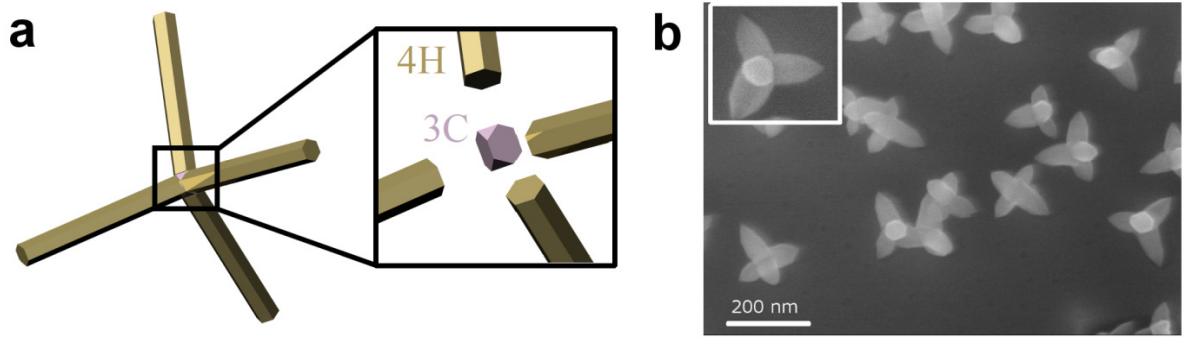

**Figure 1**. Silicon Carbide (SiC) Tetrapods. (a) Schematics of the tetrapod structure with a 3C core and 4H legs. (b) Scanning electron micrograph of the grown tetrapods. Every tetrapod has an average leg length of ~100 nm.

**Results and Discussion**

**Modeling of the tetrapod band structure**

The possible band structure in SiC tetrapods is analyzed by means of quantum mechanical simulations on a simplified model of SiC tetrapods. In this simplified model we assume a perfect and abrupt interface between 4H and 3C SiC regions, and two 4H SiC legs embedding the 3C core are taken to describe the electronic structure of the system (Figure 2a). In the other two dimensions we assume that some nearly constant profile, e.g. the cylindrical equivalent of the Bloch states of perfect 4H and 3C crystals are the solution. This treatment was already justified for similar tetrapod geometries[24-28]. As the diameter of 4H SiC legs is relatively large close to 3C core region, the dielectric confinement might minutely alter the conduction band or valence band edges of the bulk 4H SiC, so this effect is neglected. The optical excitation of the system creates a hole in the valence band and an electron in the conduction band of this system. The hot electron and hole can rapidly relax to the lowest energy excited state with the help of phonons where the electron and hole stay at the conduction and valence band edges forming an exciton. This exciton is confined along the quasi one-dimensional potential curve created by 4H (leg) - 3C (core) - 4H (leg) structure. The radiative decay of this exciton gives rise to an emission (supplementary information, Figure S1).

The quasi-one-dimensional modelling of quantum confinement of the SiC tetrapods is valid since the calculated wavefunctions of the hole and the electron resemble well the results of more elaborated 3D models of similar systems[25, 29, 30]. The key feature of these systems is the overlapping of the electron-hole pair in the quantum dot region (in the core of the tetrapod in our case), which is well described using the one dimensional model. Second, the nanometer size of the relevant quantum dot region and the bound nature of the exciton, ensure that the system's behavior is influenced by the boundary of the tetrapod.

We find that the emission wavelength of the individual tetrapods depends on their geometry, particularly, on their global symmetry, and much less on the diameter of 3C core (Figure 2b). If the length of 4H legs was about the same in a given tetrapod (global $T_d$ symmetry) then a classical rectangular quantum well forms for the electrons in the conduction band (0.92 eV) and a minor potential barrier for the holes in the valence band (0.05 eV), so that the potential curve does not show any steepness ($\Delta V = 0$). The excited electron in the conduction band is strongly confined in 3C core, and will attract the hole in the valence band to stay at 3C core region. The calculated Coulomb-energy between the electron and hole is about 0.5 eV which can surmount the small potential barrier energy for the hole in the valence band edge. Similar effects have been observed in other tetrapod systems[24-26]. Thus, both the electron and hole are confined in 3C core with emission at the proximity of 600 nm. In addition, the symmetric tetrapods should show no polarization of light. However, if the length of the 4H legs was different (i.e. not all the legs have the same length) in a SiC tetrapod then this induces different polarization of surface charges at the end of 4H legs in these tetrapods, so a steep potential curve ($\Delta V > 0$) both for the electrons and holes in the conduction and valence band edges induced by a spontaneous polarization in 4H SiC legs[31, 32] (Figure 2a). This effect creates a triangular potential well for the electron in the conduction band, and starts to push the hole away from 3C region. The exciton, which is not only bound by the Coulomb attraction of its part, but the deep triangular quantum well potential, is stable at room temperature in these tetrapods, and gives rise to emission in the range of ~ 600 – 800 nm depending on the geometry.

We also find that $\Delta V > 0.15$ eV is sufficient to separate the localization of the electron (3C core) and hole (4H leg) in these tetrapods which leads to three main differentiations with respect to the ideal case of symmetrical tetrapods: (i) longer emission wavelengths as seen from the simulation in Figure 2b), (ii) polarization of the emitted light along one of the four 4H legs, (iii) decreasing binding energy of excitons. We note that the calculated binding energies of the exciton always imply stability at room temperature, even for tetrapods with long emission wavelengths.

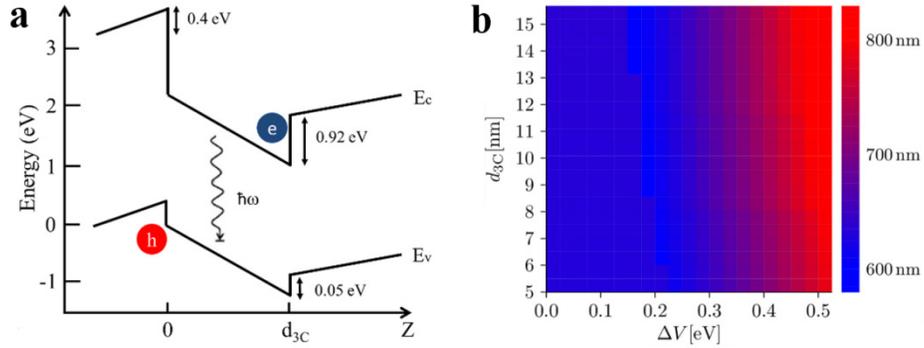

**Figure 2**. Modeling of the tetrapods. (a) Positions of hole and electron when the potential is steep in the core region. Reachable potential differences are greater than Coulomb attraction, which causes the exciton annihilation lower and the emitted photon red shifted. 0 eV was aligned to the VBM of 4H-SiC at the end of 4H-SiC legs. The band gap is 3.23 eV and 2.39 eV for the 4H-SiC and 3C SiC, respectively. (b). Calculated emission wavelengths (false color) from this model where ΔV is the potential difference along the legs and $d_{3C}$ is the diameter of the 3C core, respectively. The steep potential curve with a height of ΔV may arise from the different sizes of 4H SiC legs of a tetrapod.

**Spectroscopy**

To verify the proposed model, high resolution confocal microscopy with a 532 nm excitation (average power 200 µW) laser was employed to characterize the tetrapods. The laser was focused through a high numerical aperture objective (0.95,x100 Nikon), and the emission was collected through the same objective. The residual laser light was filtered using a dichroic mirror and a set of bandpgass filters (Semrock) and the transmitted light was focused into an optical fiber. The core of the fiber served as a confocal aperture. Figure 3a shows a confocal map of the grown tetrapods. Many fluorescent spots corresponding to single tetrapods are visible. Figure 3b shows a representative photoluminescence (PL) spectrum from a single tetrapod, revealing a very narrow and bright PL line centered at ~ 728 nm as confirmed by our simulations. The full width at half maximum (FWHM) of most of the found tetrapods is approximately ~5 nm, making them the narrowest known emitting semiconductor nanocrystals at room temperature. We imaged more than 100 tetrapods, and all exhibited similar narrow PL lines. The majority of the tetrapods exhibited fluorescence above 700 nm (supplementary information, Figure S2), in accord with the theoretical model. The majority of the tetrapods show asymmetry in their length of legs as inferred from high resolution transmision electron microscope images (supplementary information, Figure S3), that further supports the theoretical model and the flouresence distribution.

The single photon emission from the tetrapods was confirmed by photon correlation measurements using a Hanbury-Brown and Twiss interferometer. Figure 3c shows a second order correlation function, $g^{(2)}(\tau)=<I(t)I(t+\tau)>/<I(t)>^2$, recorded from a single tetrapod at minimum and at saturation excitation power. The pronounced antibunching dip in the photon statistics at zero delay time ($g^{(2)}(0)\sim 0.2$) indicates that the emission originates from a single photon emitter. The deviation from 0 is attributed to the overall background from other excitonic transitions and broad substrate luminescence (solgel). At higher excitation powers, moderate bunching behavior is observed, indicating the presence of shelving states. The whole system can then be described as a three level system, a typical model for fluorescent quantum dots[33]. At saturation we measured up to 450,000 photons/s from a single tetrapod, comparable to other single emitters in nanodiamonds and defects in bulk silicon carbide. (supplementary information, Figure S2).

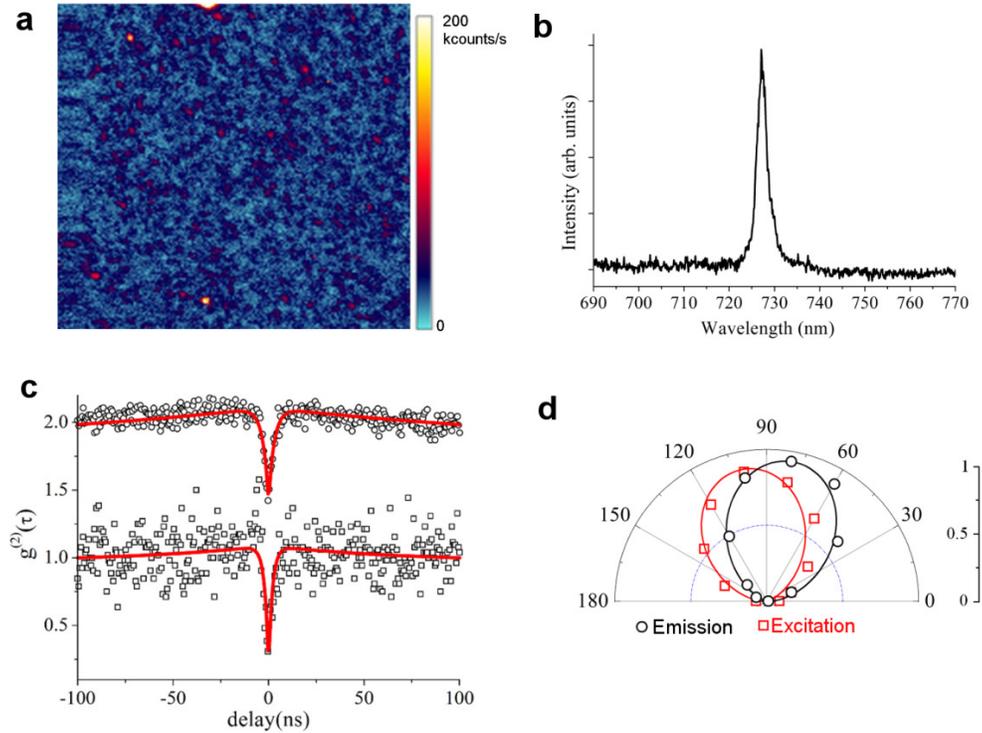

**Figure 3**. Single photon emission and polarization measurements of single tetrapods. (a) 22x22 µm$^2$ confocal map showing several tetrapods. (b) Representative PL spectrum of a single tetrapod showing with a FWHM of ~ 5 nm. (c) Second order correlation function, g$^{(2)}(\tau)$, obtained by collecting the light emitted from a single tetrapod at minimum and saturation excitation power, determines that the tetrapods are single photon emitters. The curves are displaced for clarity. (d) excitation polarization (red squares) and emission polarization (black circles) of a single tetrapod. The lines are the fit to the experimental data.

Single photon emission was reported from colloidal CdSe/Cds quantum dots embedded in rods[34] and recently from GaN quantum dots embedded in nanowires[35]. The observation of quantum emission from SiC tetrapods at room temperature is supported by a quantum-confined exciton model. The quantum confinement occurs due to the interface between 3C/4H of the SiC – forming a crystal phase quantum dot with a binding energy larger than 50 meV in any configurations of 4H SiC legs, and therefore stable at room temperature[36].

It is important to note that only a minor fraction (~10%) of the tetrapods exhibited permanent bleaching while most of them were photostable (under standard excitation conditions (~ 200 µW laser power, hours of continuous irradiation). Some of the tetrapods exhibited switching to a different state, as was reported elsewhere[23]. The photostability is another prime advantage of the tetrapods over the traditional blinking/bleaching quantum dots, where core shell structure is needed[34]. The relative stability of the tetrapods may be explained by the shielding of the exciton by the thin amorphous sheath surrounding the internal crystals in the legs and the core.

The band structure analysis predicts that the tetrapods at the longer wavelengths should exhibit fully polarized excitation and emission profiles. Motivated by the results of the simulations, polarization measurements were performed. Figure 3d show an excitation and emission polarization plots recorded from a single tetrapod. The polarization excitation measurements were carried out by positioning a half waveplate at the excitation path, while the emission polarization was measured by introducing a polarizer at the collection path.

The tetrapods exhibit very high polarization visibility, defined as $V=(I_{max}-I_{min})/(I_{max}+I_{min})$, of more than 90% at excitation and emission. The offset in the angle between excitation and emission polarization is due to repopulation of higher energy states within the tetrapod, caused by non resonant excitation. The majority of the tetrapods exhibited fully polarized excitation, as expected from a single emitting dipole and confirmed by our single photon emission measurements. Similar polarization behavior was also observed for CdS tetrapods containing CdSe quantum dots[24]. We identified only a few tetrapods emitting at the spectral range close to 600 nm that do not exhibit polarization. This is fully consistent with the prediction of the quantum confinement model.

Finally, we studied the tetrapods emission at cryogenic temperatures (4K). As can be seen from Figure 4a, several tetrapods can be observed with narrow emission lines with similar FWHM at 4K as the room temperature tetrapods. Only a few tetrapods exhibited narrowing in their FWHM down to ~ 2 nm (Figure 4b), while on average FWHM of ~ 3 nm was maintained (spectrometer resolution is below 0.1 nm). Detailed resonant excitation of a single tetrapod will be required to determine the natural linewidth of these systems. Figure 4b also shows a single tetrapod with two orthogonal polarization states, confirming the full polarization is maintained at low temperature. Full extinction of the fluorescence was not possible due to lack of alignment between the excitation field and the tetrapod dipole absorption.

The low temperature results support the hypothesis that the origin of the emission is a confined quantum system, and not a point defect. Emission from a point defect is always influenced by phonons, and the FWHM is temperature dependent and often reduced at 4K, towards Fourier transformed limited emission[37, 38]. Furthermore, as was shown recently, defects in SiC that emit single photons are extremely broad at room temperature, as typical for vacancy related defects[7]. Although in this case nanostructures rather than bulk are investigated, defect type emission is similar in both nanostructures and bulk materials[39]. Finally, the emission from 3C SiC quantum wells was also reported to have FWHM below 10 nm, at low temperature[31].

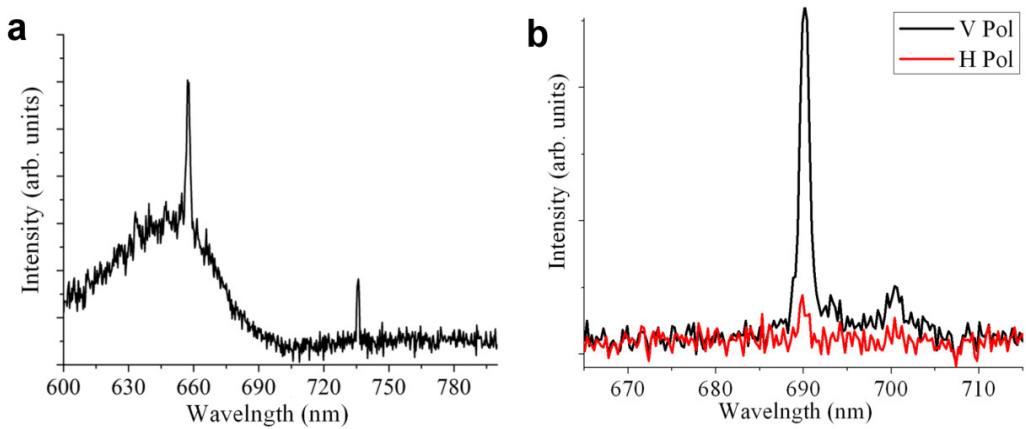

**Figure 4**. Low temperature PL of SiC tetrapods. (a) Ensemble measurement from several tetrapods, exhibiting narrow PL lines. (b) High resolution PL of a single tetrapod exhibiting fully polarized excitation profile.

**Conclusions**

To summarize, we identified a new semiconductor nanostructure, a silicon carbide tetrapod, which exhibits narrowband, bright fluorescence and single photon emission at room temperature. A systematic theoretical model confirms that the florescence originates from a quantum confinement effect between the 3C core and the 4H legs of the tetrapod. This previously unknown quantum system has the potential to be a constituent in a myriad of sensing and quantum optical applications. For instance, it can be used for high resolution bio-labeling with an additional degree of quantum imaging. Through controlled variation of the tetrapod geometry, the band structure will be altered and the emission resonance will be tuned – a highly sought after requirement for solar cells and photocatalysis[40, 41]. One can therefore select a particular emission wavelength and put together a colorful array of "quantum fluorescent labels". Such control of emission wavelengths is not possible with other room temperature single photon emitting systems. The tetrapods can also be used for quantum communications and incorporated into SiC micro electro-mechanical systems bridging the traditional microelectronic applications with quantum technologies. Finally, the SiC crystal phase tetrapods can provide an example for new nanotechnological pathways towards a new class of energy efficient nanomaterials that can deliver light harvesting significantly more efficiently than their individual counterparts.

**Acknowledgments**

The authors thank Evelyn Hu and Mor Baram for useful discussions. Dr Aharonovich is the recipient of an Australian research council discovery early career research award (project number DE130100592). AG acknowledges the support from EU commission (FP7 project No. 270197 - DIAMANT) and the Hungarian Scientific Research Fund (OTKA K101819 and K106114 projects). Part of this work was performed at the Center for Nanoscale Systems (CNS), a member of the National Nanotechnology Infrastructure Network (NNIN), which is supported by the National Science Foundation under award no. ECS-0335765. CNS is part of Harvard University. A.G. and Z.B. are also grateful for the support of the "Lendület" program of the Hungarian Academy of Sciences.

References

1. T. Basche, W. E. Moerner, M. Orrit and H. Talon, *Physical Review Letters*, 1992, **69**, 1516-1519.
2. I. Aharonovich, S. Castelletto, D. A. Simpson, C. H. Su, A. D. Greentree and S. Prawer, *Reports on Progress in Physics*, 2011, **74**, 076501.
3. A. J. Morfa, B. C. Gibson, M. Karg, T. J. Karle, A. D. Greentree, P. Mulvaney and S. Tomljenovic-Hanic, *Nano Letters*, 2012, **12**, 949-954.
4. P. Michler, A. Kiraz, C. Becher, W. V. Schoenfeld, P. M. Petroff, L. D. Zhang, E. Hu and A. Imamoglu, *Science*, 2000, **290**, 2282-+.
5. K. Hennessy, A. Badolato, M. Winger, D. Gerace, M. Atature, S. Gulde, S. Falt, E. L. Hu and A. Imamoglu, *Nature*, 2007, **445**, 896-899.
6. B. Ellis, M. A. Mayer, G. Shambat, T. Sarmiento, J. Harris, E. E. Haller and J. Vuckovic, *Nature Photonics*, 2011, **5**, 297-300.
7. S. Castelletto, B. C. Johnson, V. Ivády, N. Stavrias, T. Umeda, A. Gali and T. Ohshima, *Nature Materials*, 2014, **13**, 151-156.

8. N. Akopian, G. Patriarche, L. Liu, J. C. Harmand and V. Zwiller, *Nano Letters*, 2010, **10**, 1198-1201.
9. S. Lehmann, J. Wallentin, D. Jacobsson, K. Deppert and K. A. Dick, *Nano Letters*, 2013, **13**, 4099-4105.
10. P. Corfdir, B. Van Hattem, E. Uccelli, S. Conesa-Boj, P. Lefebvre, A. Fontcuberta i Morral and R. T. Phillips, *Nanoletters*, 2013.
11. L. Zhang, J.-W. Luo, A. Zunger, N. Akopian, V. Zwiller and J.-C. Harmand, *Nano Letters*, 2010, **10**, 4055-4060.
12. M. H. M. van Weert, N. Akopian, F. Kelkensberg, U. Perinetti, M. P. van Kouwen, J. G. Rivas, M. T. Borgstrom, R. E. Algra, M. A. Verheijen, E. Bakkers, L. P. Kouwenhoven and V. Zwiller, *Small*, 2009, **5**, 2134-2138.
13. Z. Wang, B. Tian, M. Paladugu, M. Pantouvaki, N. Le Thomas, C. Merckling, W. Guo, J. Dekoster, J. Van Campenhout, P. Absil and D. Van Thourhout, *Nano Letters*, 2013.
14. F. Fuchs, V. A. Soltamov, S. Väth, P. G. Baranov, E. N. Mokhov, G. V. Astakhov and V. Dyakonov, *Sci. Rep.*, 2013, **3**.
15. C. A. Zorman and R. J. Parro, *Physica Status Solidi B-Basic Solid State Physics*, 2008, **245**, 1404-1424.
16. A. L. Falk, B. B. Buckley, G. Calusine, W. F. Koehl, V. V. Dobrovitski, A. Politi, C. A. Zorman, P. X. L. Feng and D. D. Awschalom, *Nat Commun*, 2013, **4**, 1819.
17. P. G. Baranov, A. P. Bundakova, A. A. Soltamova, S. B. Orlinskii, I. V. Borovykh, R. Zondervan, R. Verberk and J. Schmidt, *Physical Review B*, 2011, **83**, 125203.
18. R. Krahne, G. Chilla, C. Schüller, L. Carbone, S. Kudera, G. Mannarini, L. Manna, D. Heitmann and R. Cingolani, *Nano Letters*, 2006, **6**, 478-482.
19. D. Tarì, M. De Giorgi, F. D. Sala, L. Carbone, R. Krahne, L. Manna, R. Cingolani, S. Kudera and W. J. Parak, *Applied Physics Letters*, 2005, **87**, -.
20. L. Manna, D. J. Milliron, A. Meisel, E. C. Scher and A. P. Alivisatos, *Nat Mater*, 2003, **2**, 382-385.
21. A. Fiore, R. Mastria, M. G. Lupo, G. Lanzani, C. Giannini, E. Carlino, G. Morello, M. De Giorgi, Y. Li, R. Cingolani and L. Manna, *Journal of the American Chemical Society*, 2009, **131**, 2274-2282.
22. D. V. Talapin, J. H. Nelson, E. V. Shevchenko, S. Aloni, B. Sadtler and A. P. Alivisatos, *Nano Letters*, 2007, **7**, 2951-2959.
23. A. P. Magyar, I. Aharonovich, M. Baram and E. L. Hu, *Nano Letters*, 2013, **13**, 1210-1215.
24. C. Mauser, E. Da Como, J. Baldauf, A. L. Rogach, J. Huang, D. V. Talapin and J. Feldmann, *Physical Review B*, 2010, **82**, 081306.
25. C. Mauser, T. Limmer, E. Da Como, K. Becker, A. L. Rogach, J. Feldmann and D. V. Talapin, *Physical Review B*, 2008, **77**, 153303.
26. M. C. Newton and P. A. Warburton, *Materials Today*, 2007, **10**, 50-54.
27. S. Rackauskas, K. Mustonen, T. Järvinen, M. Mattila, O. Klimova, H. Jiang, O. Tolochko, H. Lipsanen, E. I. Kauppinen and A. G. Nasibulin, *Nanotechnology*, 2012, **23**, 095502.
28. K. Sakoda, Y. Yao, T. Kuroda, D. N. Dirin and R. B. Vasiliev, *Optical Materials Express*, 2011, **1**, 379-390.

29. M. Saba, S. Minniberger, F. Quochi, J. Roither, M. Marceddu, A. Gocalinska, M. V. Kovalenko, D. V. Talapin, W. Heiss, A. Mura and G. Bongiovanni, *Advanced Materials*, 2009, **21**, 4942-4946.
30. J. Müller, J. M. Lupton, P. G. Lagoudakis, F. Schindler, R. Koeppe, A. L. Rogach, J. Feldmann, D. V. Talapin and H. Weller, *Nano Letters*, 2005, **5**, 2044-2049.
31. S. Bai, R. P. Devaty, W. J. Choyke, U. Kaiser, G. Wagner and M. F. MacMillan, *Applied Physics Letters*, 2003, **83**, 3171-3173.
32. A. Fissel, U. Kaiser, B. Schröter, W. Richter and F. Bechstedt, *Applied Surface Science*, 2001, **184**, 37-42.
33. C. Santori, M. Pelton, G. Solomon, Y. Dale and E. Yamamoto, *Physical Review Letters*, 2001, **86**, 1502-1505.
34. F. Pisanello, L. Martiradonna, G. Leménager, P. Spinicelli, A. Fiore, L. Manna, J.-P. Hermier, R. Cingolani, E. Giacobino, M. De Vittorio and A. Bramati, *Applied Physics Letters*, 2010, **96**, -.
35. M. J. Holmes, K. Choi, S. Kako, M. Arita and Y. Arakawa, *Nano Letters*, 2014, **14**, 982-986.
36. M. Dvorak, S.-H. Wei and Z. Wu, *Physical Review Letters*, 2013, **110**, 016402.
37. A. Batalov, V. Jacques, F. Kaiser, P. Siyushev, P. Neumann, L. J. Rogers, R. L. McMurtrie, N. B. Manson, F. Jelezko and J. Wrachtrup, *Physical Review Letters*, 2009, **102**, 195506.
38. T. Müller, C. Hepp, B. Pingault, E. Neu, S. Gsell, M. Schreck, H. Sternschulte, D. Steinmüller-Nethl, C. Becher and M. Atatüre, *Nat Commun*, 2014, **5**.
39. N. Mohan, Y. K. Tzeng, L. Yang, Y. Y. Chen, Y. Y. Hui, C. Y. Fang and H. C. Chang, *Advanced Materials*, 2009, **22**, 843.
40. W. Zhou, L. Yan, Y. Wang and Y. Zhang, *Applied Physics Letters*, 2006, **89**, -.
41. H. Lee, S. Kim, W.-S. Chung, K. Kim and D. Kim, *Solar Energy Materials and Solar Cells*, 2011, **95**, 446-452.